# Validation of a Conceptual Assessment Tool in E&M II

Qing X. Ryan*, Cecilia Astolfi†, Charles Baily† and Steven J. Pollock*

*Department of Physics, University of Colorado, Boulder, CO 80309-0390 USA
† School of Physics and Astronomy, University of St. Andrews, St Andrews, Fife KY16 9SS Scotland, UK

**Abstract:** As part of an ongoing project to investigate student learning in upper-division electrodynamics (E&M II), the PER research group at the University of Colorado Boulder has developed a tool to assess student conceptual understanding: the CURrENT (Colorado UppeR-division ElectrodyNamics Test). The result is an open-ended post-test diagnostic with 6 multi-part questions, an optional 3-question pretest, and an accompanying grading rubric. This instrument is motivated in part by our faculty-consensus learning goals, and is intended to help measure the effectiveness of transformed pedagogy. In addition, it provides insights into student thinking and student difficulties in the covered topical areas. In this paper, we present preliminary measures of the validity and reliability of the instrument and scoring rubric. These include expert validation and student interviews, inter-rater reliability measures, and classical test statistics.

**Keywords:** physics education research, upper-division electrodynamics, course transformation, assessment

**PACS:** 01.40.Fk, 01.40.Di, 01.40.G-, 01.40.gf

## INTRODUCTION

Research-based conceptual assessments play an important role in physics education research (PER). Such instruments are used to characterize common and persistent student difficulties, as well as to support curricular transformation. Most efforts to improve student learning have historically been aimed at lower-division courses, and multiple conceptual assessments have been developed and rigorously tested [1]. Fewer assessments are available at the upper-division level [2-4] for several reasons, including: the content is more mathematically complicated, complex upper-division physics may be perceived as less amenable to short questions, and there is a smaller pre-existing research base to inform their development.

At the University of Colorado (CU Boulder), as part of a multiyear transformation of upper-division physics courses, we developed faculty consensus learning goals, identified student difficulties and designed transformed teaching materials (including clicker questions and tutorials) for our electromagnetism (E&M) sequence [5]. We also developed conceptual assessments [4, 6] to measure those learning goals that are typically not directly addressed/tested by traditional exams. In particular, our own traditional exams generally target calculational skills, while the assessments are largely focused on conceptual underpinnings. Such an assessment serves different purposes for different stakeholders. It allows for comparisons across different academic years and different institutions, and provides insights for instructors about students' conceptual difficulties in core topics. Students can use this test as a review to help them prepare for the final exam. Research-based assessment is informative to the PER community as a tool to better understand common student difficulties, to inform curriculum development, and to assess pedagogical reforms.

The *Colorado UppeR-division ElectrodyNamics Test* (CURrENT) [5, 6], guided by our course-scale faculty consensus learning goals [5], is designed to measure a representative sampling of skills and conceptual understanding in selected core topics from the *second* semester of E&M, covering electrodynamics (E&M II: Griffiths [7] Ch.7-12). Note that a separate instrument, the CUE, measures student achievement of learning goals in statics (E&M I: Griffiths Ch. 1-6) [4]. In this paper, we summarize the ongoing development of the CURrENT instrument, and present preliminary measures of its validity and reliability with data across 6 different institutions and 271 students (40% of the students are from CU Boulder).

## OVERVIEW

The CURrENT has 6 multi-part questions, with 15 sub-questions, which further break down into a total of 47 scoring elements (the smallest check point where students get a score). The open-ended format requires students to justify their answers by also providing reasoning. The post-test requires about 50 minutes. The pretest (20 min.) contains a subset of questions that have been modified in order to be accessible to students who have not yet seen the material. A brief description of each question, along with the learning goals targeted, is given in Table 1.

TABLE 1. Summary of items on the CURrENT, including point allocations, learning goal alignment, and Cohen's κ (N=90)(see section on reliability). Question numbers in bold are also on the pretest. The targeted learning goals are: 1. Math/physics connection 2. Visualize the problem 4. Communication 5. Problem-solving techniques 6. Problem-solving strategy 7. Expecting/checking solution 9. Maxwell's Equations 10. Build on earlier material. For more detail, and content-specific goals, see reference [5].

| Q #. | Pt | Description | Goals | κ |
|---|---|---|---|---|
| **Q1a** | 5 | Integral form of Maxwell eqns. | 1,9 | 0.95 |
| **Q1b** | 5 | Integral elements and visualization | 1,2,4,5 | 0.70 |
| **Q2a** | 5 | B field of ∞ solenoid | 2,7,10 | 0.82 |
| Q2b | 5 | E field of time-varying solenoid | 2,5,6,10 | 0.91 |
| **Q3a** | 5 | Math expression of an integral statement | 1 | 0.85 |
| **Q3b** | 5 | Vector calculus derivation | 5 | 0.78 |
| Q4a | 5 | E field continuity | 4 | 0.91 |
| Q4b | 5 | Divergence of J | 1,4 | 0.90 |
| Q5a | 5 | Energy density of charging capacitor | 1,4 | 0.88 |
| Q5b | 5 | Poynting vector | 1,4,7 | 0.77 |
| Q6a | 2 | \|E\| in complex notation | 1,4 | 1 |
| Q6b | 2 | Index of refraction | 4,10 | 1 |
| Q6c | 2 | E at boundary | 1,2,4,9 | 1 |
| Q6d | 2 | B at boundary | 1,2,4,9 | 0.97 |
| Q6e | 2 | Continuity@boundary | 1,4,5 | 0.94 |

## VALIDITY

Validity is defined as the extent to which test scores accurately measure the intended concept or construct. Does the instrument give similar results to other approaches that measure the same construct? Do experts agree with the way these constructs were operationalized to achieve the corresponding learning goals? Do students interpret the questions as intended?

**Criterion Validity:** One indicator of validity is the extent to which the test gives results similar to other independent measures. CURrENT scores correlate highly with students' final exams in their junior E&M course (Pearson correlation coefficient $r=0.52$, $p<0.001$, $N=271$) and their course grades ($r=0.46$, $p<0.001$, $N=271$). These correlations are considered ''medium'' (0.3–0.5) to ''strong'' (0.5–1.0) [8], suggesting that the constructs measured on the CURrENT are highly related to other aspects of student performance typically valued by faculty.

**Expert Validity:** We would like to know if experts agree that the assessment questions measure the course-scale learning goals. We surveyed seven experienced faculty members (two from research institutions and five from 4-year colleges) and one PER researcher by providing them with the CURrENT and the corresponding course scale learning goals matched to each question. Faculty were asked to provide feedback via two guiding questions: 1) Do you think these questions are something you expect your students to be able to answer? 2) Is there anything on this list (learning goals aligned to each question) that appears to be inappropriate or mismatched to you?

All eight stated that they found the questions to be valuable and useful. Four experts proposed small adjustments of which learning goals best aligned with the questions, and these adjustments have been incorporated into our final list. Two faculty members have a different course structure where one junior-level module covers both electrostatics and electrodynamics. We included their opinions to assess the relevancy of the assessment in a broader sense. Both expressed approval for the test content, describing it as "interesting" and "a good cross section" of the material expected to be covered in the EM course. One faculty member thought the sub-question Q3b that addresses the derivation learning goal (sub-goal of goal 5) was unimportant to them. However, "derivation and proof" is an important learning goal at CU and it is not surprising that different instructors have different emphasis on learning goals at this level. Our goal in matching up questions with specific learning goals was to help instructors understand what the assessment is trying to measure, and whether it is aligned with their own learning goals

**Student Validation:** In addition to numerous student interviews conducted during the development process, we validated the most recent version to determine whether students are interpreting the questions as we intended. We interviewed four students using a think-aloud protocol: the interviewer did not interject except to remind students to verbalize their thought processes. The interviews were recorded and later analyzed to determine whether student work reflected the intended nature of the question, as well as whether their written work reflected their verbal interpretation of the question. At the end of the interview, students were asked about questions that seemed problematic, to probe their understanding of the question prompt. A few wording and spacing changes were made as a result of these student interviews. For example, Q1b was originally worded as: *If this surface cannot be used with Eq. I or Eq.II, briefly explain why not in the space below.* Almost all students knew that the given surface can be used with the second equation, so we modified this question slightly because students suggested the wording was confusing, and weren't sure if an explanation should be provided for both equations.

# RELIABILITY

Reliability is defined as the overall consistency and stability of a test measure. We concentrate on two aspects of reliability: 1) Does students performance on any given test item correlate with the remaining items on the test (internal consistency)? 2) How well do different scorers agree with each other on the same set of students (inter-rater reliability)?

**Internal Consistency:** Cronbach's alpha ($\alpha$) is a statistical measure of internal consistency. We treat each sub question (e.g. 1a, 1b) as a single test item, and obtained $\alpha=0.72$ (N=271), where $\alpha$-values between 0.7-0.9 are traditionally considered adequate [9]. We also computed $\alpha$ more conservatively by treating each question (including all sub parts) as one test item and obtained $\alpha=0.69$. Cronbach's $\alpha$ assumes unidimensionality of test items. We have no evidence that the CURrENT measures a single construct, suggesting that values of $\alpha$ are likely a conservative underestimate of internal consistency [10, 11].

**Inter-rater Reliability:** When designing a detailed scoring rubric for the CURrENT, we tried to make the grading objective and straightforward, so as to achieve high reliability while requiring as little training as possible. To check inter-rater reliability, 90 CURrENT exams (consisting of two sets: CU (N=47) and a 4-yr college (N=43)) were scored independently by two different raters. Rater 1 was a PER faculty (SJP) and rater 2 was a PER postdoc researcher (QR) who had not scored the test previously. Raters discussed all questions for 11 exams randomly selected from the CU set as an initial training, and then scored the rest independently. After the independent scoring, raters discussed only those 8 (out of 47) scoring elements where agreement fell below 90% for the non-CU set. Data reported below includes the initial training set.

The inter-rater reliability for the two raters working independently was very high: the total CURrENT score given by these two raters only differed by 0.2% (0.01% after discussion). We also looked at the absolute value of rater difference on the "total CURrENT score" to provide a spectrum of the variation on individual students. The absolute value of rater difference has an average of 3% (1% after discussion) and a standard deviation of 3% (2% after discussion). Raters agreed on overall scores to within ±5% for the vast majority of the students (86%). Raters differed by more than 10% on only 2% of the students (N=2) and the CURrENT score after discussion differed by 0% and 2% respectively for these two students.

We also examined inter-rater reliability on individual questions. Rater difference (absolute value) on any individual question, averaged 3.5% (1.0% after discussion) and the standard deviation of differences was 9.5% (4.5% after discussion). Similar results were found with a third rater (an undergraduate researcher, CA), suggesting that these findings are generalizable.

As an additional measure of inter-rater reliability, we computed Cohen's kappa ($\kappa$) [12]. This is a statistical measure which indicates how often raters give an exam, or a question, the same score, compared to the proportion expected by chance. We generated a contingency table based on all possible scoring combinations and computed $\kappa$ (N=90) for all sub-questions. All of our $\kappa$ values (raters 1&2) are "substantial" or better [13], suggesting a satisfactory inter-rater agreement (See Table 1).

# DISCRIMINATION

Lastly, we offer some measures as to how well the test discriminates students with different abilities. We would like to see consistent discrimination across questions, a broad distribution of total scores, and a reasonable level of difficulty of test items.

**Item-test Correlation:** We expect that students who score well on the test as a whole will tend to score well on individual items. The item-test correlation is typically calculated in terms of point-biserial correlation, but this is only applicable for dichotomous variables. For this open-ended test format, we instead examined the Pearson correlation coefficient for each test item with the rest of the test (with the item itself excluded). The correlation coefficient for each of the six questions ranged from 0.40 and 0.49. Minimum acceptable correlation coefficients are generally considered to be around 0.2 [14]. The overall distribution of the CURrENT scores is shown in Fig. 1, which also visually indicates the discriminatory power of the test (good distribution across all score bins) with the normality of the data verified by an Anderson-Darling test (p>0.05).

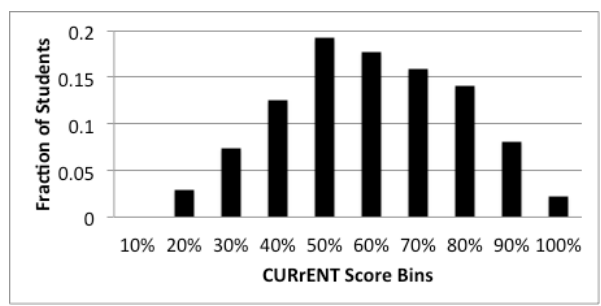

**FIGURE 1. Histogram of** CURrENT scores (N=271).

**Coefficient of Test Discrimination:** Ferguson's delta ($\delta$), or the "coefficient of test discrimination" [15], measures the discriminatory power of a test by investigating how broadly the total scores of a sample are distributed over the possible range [14].

Calculating Fergusons' δ for a multiple-choice test (e.g. [14]) is straightforward because the number of items and the possible score bins are unambiguous. However, there is not a well-accepted method for calculating δ for open-ended assessments. We used two reasonable alternatives: 1) take the total number of test items (K) as the number of points on the test, and calculate the frequency ($f_i$) of the number of points earned [4]; or 2) convert the open-ended scoring to multiple choice, simply turning all scoring elements to a corresponding 0 or 1. We obtained δ =0.99 with the first method, and 0.98 with the second method. The possible range of δ values is [0,1]. Traditionally, δ > 0.9 is considered good discrimination and thus the CURrENT has substantial discrimination power in differentiating students with different abilities.

**Item Difficulty:** The item difficulty index [14] statistic is not applicable because the open-ended scoring is not dichotomous. Instead, we compute the mean for each question to give an idea of how difficult each item is. As shown in Fig. 2, different questions on the CURrENT pose different levels of challenge for students, with additional variation evident across student populations. None of the questions yield an extremely high or low percentage, indicating an appropriate level of difficulty for purposes of discrimination. Further, despite the differences between classes/institutions, students across these institutions scored consistently lower on some questions than others (for example, students in 3 courses scored consistently lower on Q2, and those in 2 courses scored lower on Q4 and Q5). In other words, some questions are systematically more difficult across different populations, indicating common students difficulties that should be investigated and addressed.

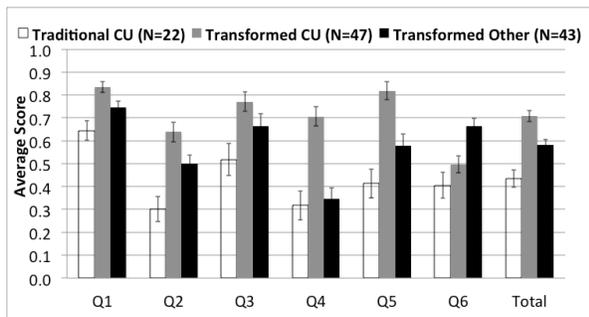

**FIGURE 2.** CURrENT results of three different courses for individual questions and total score. Error bars represent the standard error of the mean.

## CONCLUSIONS

The validity and reliability of the CURrENT has been evaluated. Validation results with experts and students were overall positive, yielding only small changes in wording. The CURrENT score is well-correlated with other variables, such as final exams and course grades, that are typically valued by faculty. The test shows high internal consistency and a high degree of inter-rater reliability using the accompanying rubric. We can differentiate between students with different abilities with this test, and are discerning measurable differences between different pedagogies (i.e., traditional courses or transformed courses in Fig. 2). This instrument shows considerable promise for research and assessment in upper-division electro-dynamics. Data collected from administering this test is also adding to our research base on common student difficulties. We are still in the process of improving the grading rubric, and are engaging in preliminary studies to explore the possibility of developing a multiple-choice version of the test, in order to minimize grading efforts and further eliminate subjectiveness of grading.

## ACKNOWLEDGEMENTS

We gratefully acknowledge contributions of many students and faculty including: B. Ambrose, A. Becker, B. Braunecker, S. Chasteen, P. Kohl, F. Kontur, B. Sinclair, B. Wilcox and the PER@C group. This work is supported by CU Boulder, the CU Science Education Initiative and NSF-CCLI grant #1023208.

## REFERENCES

1. D. Meltzer & R. Thornton. *AJP* **80**, 478 (1992).
2. G. Zhu & C. Singh, *AJP* **80**, 252 (2012);
3. B. Notaros (2002); see tinyurl.com/Notaros-EMCI.
4. S. Chasteen, et al., *PRST-PER* **8**, 020108 (2012).
5. For access to course materials and learning goals, see *http://per.colorado.edu/sei*
6. C. Baily, M. Dubson & S. Pollock, *PERC Proc. 2012* (AIP, Melville, NY, 2013), p.54-57.
7. D. Griffiths, *Introduction to Electrodynamics, 3rd Ed.* (Prentice-Hall, Upper-Saddle River NJ, 1999).
8. J. Cohen, *Statistical Power Analysis for the Behavioral Sciences* (Routledge, NY, '88), 2nd ed.
9. J. Nunnally, (1978). *Psychometric theory* (NY: McGraw-Hill, 1978) 2nd ed.
10. Cortina, J.M., *J. App. Psych.* **78**, 98–104 (1993).
11. R. Brennan & D. Prediger, *Educ. Psychol. Meas.* **41**, 687 (1981).
12. J. Cohen, *Psychol. Bull.* **70**, 213 (1968).
13. J. Landis and G. Koch, *Biometrics* **33**, 159 (1977).
14. L. Ding et al., *PRST-PER* **2**, 010105 (2006).
15. G. Goldstein & M. Hersen, *Handbook of Psychological Assessment,* (Kidlington, Oxford, UK, 2000), 3rd ed.